\documentclass[twocolumn,showpacs,preprintnumbers,nofootinbib,prd,
superscriptaddress,10pt ]{revtex4-2}

\usepackage{lmodern}
\usepackage{amsmath,amssymb}
\usepackage{physics}
\usepackage[normalem]{ulem}
\usepackage{textcomp}
\usepackage{hyperref}
\usepackage{bm}
\usepackage{graphicx}
\usepackage{psfrag}
\usepackage[usenames,dvipsnames]{xcolor}
\usepackage[utf8]{inputenc}
\usepackage{lipsum}
\usepackage{aas_macros}

\definecolor{darkgreen}{rgb}{0,0.5,0}

\hypersetup{
    bookmarks=true,         % show bookmarks bar?
    unicode=false,          % non-Latin characters in Acrobat???s bookmarks
    pdftoolbar=true,        % show Acrobat???s toolbar?
    pdfmenubar=true,        % show Acrobat???s menu?
    pdffitwindow=false,     % window fit to page when opened
    pdfstartview={FitH},    % fits the width of the page to the window
	pdftitle={title},
    pdfauthor={authors},
    pdfsubject={Gravitational Waves},   % subject of the document
    pdfkeywords={gravitational waves, eccentric compact binaries, general relativity},
    pdfnewwindow=true,      % links in new window
    colorlinks=true,       % false: boxed links; true: colored links
    linkcolor=red,          % color of internal links
    citecolor=Blue,        % color of links to bibliography
    filecolor=magenta,      % color of file links
    urlcolor=Blue,           % color of external links
    linktocpage=true
}

\allowdisplaybreaks[4]

\DeclareFontFamily{OT1}{pzc}{}
\DeclareFontShape{OT1}{pzc}{m}{it}{<-> s * [1.10] pzcmi7t}{}
\DeclareMathAlphabet{\mathpzc}{OT1}{pzc}{m}{it}

\usepackage{graphicx}
\usepackage{subcaption}
\graphicspath{{./figures/}}
\begin{document}

\title{Inferring additional physics through unmodelled signal reconstructions}

\date{\today}

\author{Rimo Das}
\email{rimo.physics@gmail.com}
\affiliation{Department of Physics, Indian Institute of Technology Madras, Chennai 600036, India}
\affiliation{Centre for Strings, Gravitation and Cosmology, Department of Physics, Indian Institute of Technology Madras, Chennai 600036, India}

\author{V. Gayathri}
\email{gayathri.v@ligo.org}
\affiliation{Leonard E. Parker Center for Gravitation, Cosmology, and Astrophysics, University of Wisconsin–Milwaukee, Milwaukee, Wisconsin 53201, USA}

\author{Divyajyoti}
\email{divyajyoti.physics@gmail.com}
\affiliation{Gravity Exploration Institute, School of Physics and Astronomy, Cardiff University, Cardiff, CF24 3AA, United Kingdom}
\affiliation{Department of Physics, Indian Institute of Technology Madras, Chennai 600036, India}
\affiliation{Centre for Strings, Gravitation and Cosmology, Department of Physics, Indian Institute of Technology Madras, Chennai 600036, India}

\author{Sijil Jose}
\email{sijiljose.999@gmail.com }
\affiliation{Department of Physics, Indian Institute of Technology Madras, Chennai 600036, India}

\author{Imre Bartos}
\email{imrebartos@ufl.edu}
\affiliation{Department of Physics, University of Florida, PO Box 118440, Gainesville, FL 32611-8440, USA}
\author{Sergey Klimenko}
\email{klimenko@phys.ufl.edu}
\affiliation{Department of Physics, University of Florida, PO Box 118440, Gainesville, Florida 32611-8440, USA}

\author{Chandra Kant Mishra}
\email{ckm@physics.iitm.ac.in}
\affiliation{Department of Physics, Indian Institute of Technology Madras, Chennai 600036, India}
\affiliation{Centre for Strings, Gravitation and Cosmology, Department of Physics, Indian Institute of Technology Madras, Chennai 600036, India}

\begin{abstract}
Parameter estimation of gravitational wave data is often computationally expensive, requiring simplifying assumptions such as circularisation of binary orbits. Although, if included, the subdominant effects like orbital eccentricity may provide crucial insights into the formation channels of compact binary mergers. To address these challenges, we present a pipeline strategy leveraging minimally modelled waveform reconstruction to identify the presence of eccentricity in real time. Using injections (40$\rm{M_{\odot}}$ binary black hole signals), we demonstrate that ignoring eccentricity (with values larger than $\sim$0.15 estimated at 20Hz ($e_{20}$)) leads to significant biases in parameter recovery, including chirp mass estimates falling outside the 90\% credible interval. Waveform reconstruction shows inconsistencies increase with eccentricity, and this behaviour is consistent for different mass ratios. Our method enables low-latency inferences of binary properties supporting targeted follow-up analyses and can be applied to identify any physical effect of measurable strength.
\end{abstract}

\date{\today}
\maketitle

\section{Introduction}
\label{sec:intro}
During the second part of the third observation run (O3b) LIGO~\cite{LIGOScientific:2014pky} and Virgo~\cite{VIRGO:2014yos} detectors observed compact binary mergers with a frequency of at least one per week \cite{LIGOScientific:2021djp}. In the ongoing observation run (O4), detection rates have gone up by a factor of few as expected~\cite{Shoemaker:2019bqt, McClelland:T1500290-v3, KAGRA:2013rdx} compared to those in O3b, potentially putting significant strain on the available human and computational resources. While there have been numerous proposals in the past to help cut analysis costs, for instance, by constructing signal models that are fast to generate (see \cite{Purrer:2014fza} for example) or by developing novel analysis methods (such as those of \cite{Smith:2016qas}), post-detection analyses for signals with generic features are still computationally prohibitive. As capabilities of current detectors improve (Voyager~\cite{McClelland:T1500290-v3}), precise measurements of subdominant effects, such as the presence of orbital eccentricity, spin precession and higher order modes should become possible. However, quasicircular precessing binary black hole (BBH) mergers can be mistaken for highly eccentric mergers due to the degeneracy between the two effects~\cite{CalderonBustillo:2020xms}. Currently, a few events have been reported to be eccentric, GW190521 being one of them ~\cite{Gayathri:2020coq,Romero-Shaw:2020thy,Iglesias:2022xfc}. Further, inclusion of eccentricity may even become necessary for analysing data from planned future ground-based detectors like Cosmic Explorer (CE)~\cite{Dwyer:2014fpa} and Einstein Telescope (ET)~\cite{Punturo:2010zz}, 
and space-based detectors such as DECIGO~\cite{Sato_2017} and LISA~\cite{amaro2017laser} owing to their low-frequency sensitivity.\\
\\
Parameter estimation (PE) of compact binary mergers is typically performed over a 15-dimensional parameter space comprising intrinsic parameters such as component masses and spins and a set of extrinsic parameters including those giving the location and orientation of the binary \cite{LIGOScientific:2016vlm}. Additionally, assumptions that the binary's orbits are noncircular or that its constituents are susceptible to tidal forces further extend the parameter space. However, if one can infer the absence of a specific physical effect, say for instance, of eccentricity, in that case, one can exclude it from the parameter space to be explored, thus reducing the analysis time. Conversely, observing a binary with residual orbital eccentricity in current ground-based detectors may suggest that the binary was influenced by external factors; for instance, it may have been part of a hierarchical triple system \citep[e.g.,][]{Antonini:2017ash}, was in a densely populated star cluster \citep[e.g.,][]{Rodriguez:2017pec}, or was within the accretion disk of a supermassive black hole \citep[e.g.,][]{Tagawa:2020jnc,Bartos:2016dgn,Yang:2020xyi,Gayathri:2021xwb}. Furthermore, binaries formed through dynamical interactions in dense stellar environments \citep{PortegiesZwart:1999nm, 2010MNRAS.407.1946D, Rodriguez:2016kxx, Banerjee:2017mgr, DiCarlo:2019pmf, Mapelli2020, Mandel:2018hfr, Samsing:2017xmd, Fragione:2018yrb, Samsing:2020tda} or via Kozai-Lidov mechanisms \citep{Kozai:1962zz, LIDOV1962719} in field triples \citep{Martinez:2020lzt, 2016ARA&A..54..441N} could exhibit residual eccentricities of $\gtrsim0.1$ when observed by ground-based detectors \citep[e.g.,][]{2011A&A...527A..70K, Lower:2018seu, Samsing:2017rat, Zevin:2021rtf}. However, we may miss such opportunities if eccentricity information is not included in the post-detection analyses.

\vspace{5pt}

In recent years, several efforts have been made to develop eccentric waveform models which may be available for use in upcoming observing runs for assessing the presence of eccentricity and its impact on understanding source properties. These include several inspiral-only models for gravitational wave (GW) signals from eccentric compact binary systems, which are sufficiently accurate so that they can be compared with numerical relativity (NR) simulations, and are rapid enough to generate for use in direct parameter estimation via Bayesian inference \cite{Konigsdorffer:2006zt, Yunes:2009yz, Klein:2010ti, Mishra:2015bqa, Moore:2016qxz, Tanay:2016zog, Klein:2018ybm, Boetzel:2019nfw, Ebersold:2019kdc, Moore:2019xkm, Klein:2021jtd, Khalil:2021txt, Paul:2022xfy, Henry:2023tka}. Further, eccentric waveform models containing the inspiral, merger, and ringdown (IMR) are under development and/or are available for use \citep[e.g.,][]{Hinder:2017sxy, Huerta:2017kez, Cao:2017ndf, Chiaramello:2020ehz, Islam:2021mha, Chattaraj:2022tay, Ramos-Buades:2021adz, Liu:2023dgl, Ramos-Buades:2023yhy, Paul:2024ujx,Liu:2019jpg, Hinderer:2017jcs, Albanesi:2022xge, Ramos-Buades:2021adz, Nagar:2021gss, Iglesias:2022xfc, Liu:2021pkr,Islam:2021mha, Yun:2021jnh, Huerta:2017kez}, although these are typically slower to generate compared to their quasicircular counterparts. Consequently, Bayesian inference with these models has often necessitated relaxing accuracy requirements \citep[e.g.,][]{OShea:2021faf}, employing likelihood reweighting methods \citep[e.g.,][]{Romero-Shaw:2019itr}, or relying on highly resource-intensive parallel inference performed on supercomputer clusters \citep[e.g.,][]{pBilby, Romero-Shaw:2022xko}. 

\vspace{5pt}

\begin{figure*}[t!]
    \centering
    \includegraphics[width=\textwidth]{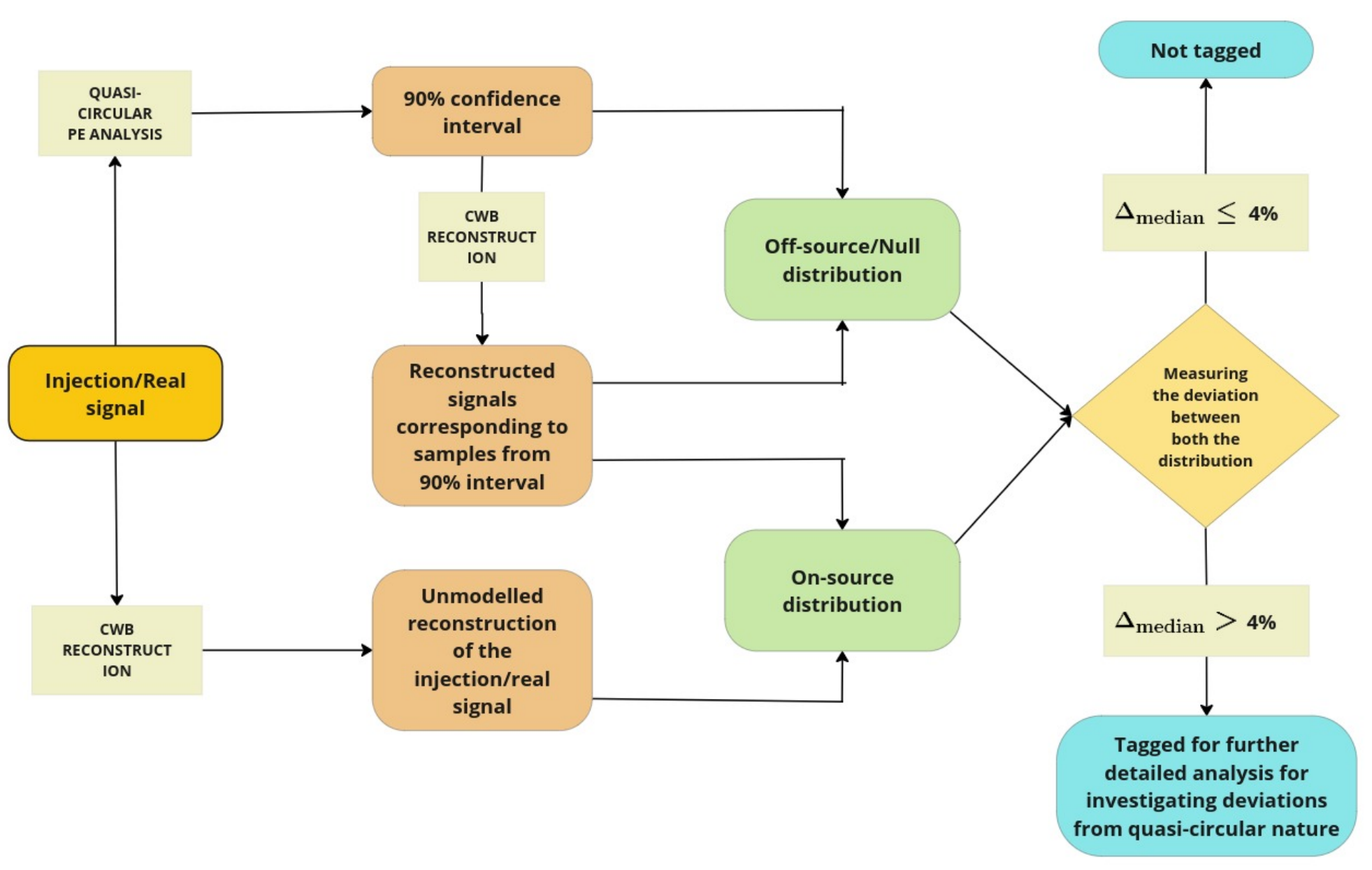}
    \caption{Flowchart describing the methodology used in this paper. The injected signal is reconstructed using cWB (lower brown block) and signals corresponding to the PE samples within 90\% credible interval (upper brown block) are reconstructed again using cWB (middle brown block). Null distribution is obtained by calculating overlap of PE samples with corresponding reconstructions. Subsequently, the injection fitting factor distribution is obtained using overlap between the reconstruction of injected waveform and reconstructions corresponding to PE samples. The deviation between the two distributions is characterised by $\Delta_{\rm{median}}$.  We choose to work with a conservative choice of 4\% for the upper bound on $\Delta_{\rm median}$ estimates as a criterion for classification of the injection as noneccentric (see Fig.~\ref{fig:median} as well as Sec.~\ref{sec:delta_median} for details).}
    
    \label{fig:method}
\end{figure*}

Numerous parameter estimation studies have been conducted to investigate the presence of orbital eccentricity in signals identified by standard searches optimized for quasicircular BBHs. These studies utilize available eccentric waveform models through Bayesian inference methods \citep[e.g.,][]{Lower:2018seu, Romero-Shaw:2019itr, Romero-Shaw:2020thy, Lenon:2020oza, Romero-Shaw:2021ual, Romero-Shaw:2022xko, Romero-Shaw:2022fbf, Iglesias:2022xfc, Gamba:2021gap, Bonino:2022hkj, Romero-Shaw:2020aaj} or compare the data directly with NR simulations of gravitational waves from eccentric BBH \citep{Gayathri:2020coq}. There have also been a number of studies \citep{Divyajyoti:2023rht, Favata:2013rwa, Abbott:2016wiq, Favata:2021vhw, Ramos-Buades:2019uvh, Narayan:2023vhm, Guo:2022ehk, Saini:2022igm, Bhat:2022amc, Bonino:2022hkj, GilChoi:2022nhs} highlighting the biases induced in parameters when eccentric signals are analysed with quasicircular waveforms. \\
\\
While the use of eccentric waveforms in template-based searches may only be realised in the future, search methods that are not sensitive to the details of the signal morphology present a suitable alternative~\cite{Klimenko:2005xv, Salemi:2019uea, Tiwari:2015gal,Roy:2022teu}.\footnote{No evidence for the presence of eccentricity was found using these methods in the data for the first two observing runs \cite{Salemi:2019owp,Romero-Shaw:2022xko}.} Moreover, since the first GW detection, GW150914 \cite{Abbott:2016blz}, waveform reconstruction methods have been essential for evaluating the consistency between unmodelled reconstructions and PE results. Tools like Coherent WaveBurst \cite{Klimenko:2005xv, 2008CQGra..25k4029K,Salemi:2019uea,Klimenko:2015ypf} (cWB) and BayesWave \cite{Cornish_2015,PhysRevD.103.044006} have played a key role in these efforts, with cWB providing constrained maximum-likelihood reconstructions and BayesWave utilizing the median from its posterior probability distribution. These methods have been applied extensively in gravitational-wave transient catalogues \cite{LIGOScientific:2018mvr,LIGOScientific:2020ibl,LIGOScientific:2021usb,KAGRA:2021vkt} and individual event analysis, such as the detailed study of GW190521 \cite{Abbott:2020tfl,Szczepanczyk:2020osv,Gayathri:2020coq}, a rare and significant event characterized by high mass and spin-precession measurements. Tests using these methods have extracted critical physics, including higher-order modes and eccentricity. While most studies have focused on real events, the potential of waveform reconstruction for simulations remains under-explored. Leveraging these techniques in the cWB framework can serve as a powerful tool to identify events requiring detailed follow-up, such as eccentric PE studies, motivating a deeper investigation into waveform reconstruction for eccentric systems.\\
\\
The current work attempts to develop a strategy that may eventually be used to infer the presence or absence of orbital eccentricity (together with other physical effects) in observed events and, therefore, guide the offline analyses that follow the detections.
Here, we target inferring the presence of orbital eccentricity in a simulated signal as a proof of principle demonstration of the method. The method, in principle, could be used to infer the presence of any additional physical effect or a superposition of subdominant effects that are typically ignored due to associated computational cost. The basic idea is as follows.\\
\\
Let us say we have a signal present in a noisy data stream. The signal may correspond to a circular or an eccentric binary, but we do not have this information \textit{a priori}. In order to infer the nature of the binary, we propose the following. First, we perform a detailed parameter estimation exercise using a circular state-of-the-art model.\footnote{In practice, we may simply use the results of any online PE analysis\citep{Nitz:2018rgo} so as to perform the exercise with low latency.} Naturally, the measurements will be consistent with a circular binary even though the signal is eccentric, and our estimates may be biased. The bias may be small or large depending upon the binary's eccentricity. In any case, the results (posterior distributions) will represent an effective measurement of source properties. If one also has an eccentric waveform that can be used for PE, one could simply repeat the exercise and obtain an unbiased estimation of parameters together with a measurement of the binary's orbital eccentricity. But let's assume we do not have such a model, or even if we do, using it for a detailed analysis is computationally expensive. In such situations, we propose that search methods developed within the cWB framework on data may be employed.  While these methods are routinely recommended for analyzing short duration signals, as we shall see in upcoming sections, these can also be faithfully used for analyzing longish ($\sim30$-40 cycles~\cite{Chattaraj:2022tay}) binary black hole signals. The output is a reconstructed signal that is expected to capture all relevant physics of the signal. One may then plan to reconstruct signals based on posterior distributions of the PE analysis and compare with the one obtained using direct reconstruction above. The expectation is that if the signal is circular, the two reconstructions will match. Hence, any significant loss of match will indicate the presence of eccentricity in the system and thus provide an indirect inference of the same. This methodology can be extended to other effects which may modify the gravitational wave signal such as compact binaries in the presence of astrophysical environments \cite{CanevaSantoro:2023aol, Leong:2023nuk}, mergers of boson star binaries \cite{Evstafyeva:2024qvp, CalderonBustillo:2022cja}, etc. but are not currently accounted for in the standard PE exercises. Figure \ref{fig:method} displays the detailed methodology. 

Our investigations involve injecting a set of eccentric IMR signals for both PE and waveform reconstruction studies. These were obtained from hybrids constructed by combining post-Newtonian (PN) waveforms and NR simulations in an earlier work \cite{Chattaraj:2022tay}.
This paper is organised as follows. Section \ref{sec:method} summarises the methodology adopted for quantifying the detection of an eccentric GW mode. Waveforms, detector configuration and other necessary inputs are also included in this section. Section \ref{sec:res} presents our findings. Finally, Sec.~\ref{sec:concl} includes a comprehensive summary of our findings and future directions.

\section{Methodology}
\label{sec:method}

\subsection{Simulations}
\label{sec:method-wf}

The full inspiral-merger-ringdown hybrids are constructed by matching the PN waveforms and NR simulations. The matching is performed in a region where the PN prescription matches the NR data by more than 99\% following the method in Ref \cite{Varma:2016dnf}. 
The PN expressions of GW modes for the inspiral part, which are 3PN accurate in amplitude~\cite{Boetzel:2019nfw, Ebersold:2019kdc, Mishra:2015bqa} and 3.5PN accurate in phase \cite{koenigsdoerffer-2006, Tanay:2016zog}, assuming nonspinning compact objects in quasielliptical orbits, are computed based on Quasi-Keplerian representation in Refs~\cite{damour-1985, Memmesheimer:2004cv, Arun:2008kb}. 
The NR simulations, which model the late inspiral and merger-ringdown phases to capture strong-field gravity effects, have been performed using the Spectral Einstein Code  developed by the SXS Collaboration \cite{Ossokine_2013, Boyle:2019kee}. In our study, we considered only ($l=2$, $|m|=2$) or simply (2,2) mode for the hybrid signal. The effects of subdominant modes such as ($l,|m|$)= (3, 3), (4, 4), (5, 5), (2, 1), (3, 2), and (4,3) have not been used in this analysis. The reason for taking only the dominant mode is that any additional systematics other than eccentricity can bias the parameter estimation study, affecting our further cWB analysis.  In our analyses, we have used twenty-one nonspinning eccentric hybrids from Ref~\cite{Chattaraj:2022tay} with mass ratios ($q$ = 1,2,3). These simulated signals act as a proxy for the actual GW signal in the detector. The details of the hybrids are mentioned in Table I of Ref \cite{Chattaraj:2022tay}.

\subsection{CWB reconstructions}
\label{sec:method-cwb}

Coherent WaveBurst (cWB) \citep{Klimenko:2005xv, 2008CQGra..25k4029K, Klimenko:2015ypf, Tiwari:2015gal}, a model-independent search pipeline has been used by the LIGO-Virgo-Kagra (LVK) Collaboration to both detect and reconstruct gravitational waveforms without assumptions about the source. cWB uses wavelet transformation to convert detector data into a time-frequency representation and identifies potential candidates by clustering time-frequency pixels with higher coherent energy. These clusters undergo a series of noise veto checks, with surviving candidates considered gravitational-wave events.

For each event, cWB computes summary statistics describing properties such as signal duration, central frequency, and correlation across detectors. Thresholds are applied to these statistics to distinguish true GW events from noise fluctuations, improving the significance of candidate events.  

A key feature of cWB is its ability to reconstruct waveforms for detected candidates through inverse wavelet transformation, a process that makes minimal assumptions about the signal shape and is often referred to as minimal modelling. By comparing these reconstructed waveforms with those obtained through PE using quasicircular templates, we can identify unexpected behaviours in GW events that are not captured by traditional PE models. In particular, cWB’s model-independent approach allows it to maintain sensitivity to eccentric mergers~\cite{Chandra:2020ccy,Bhaumik:2024cec}, making it a powerful tool for studying eccentric systems, especially since at present there is no template bank that may be used to perform an eccentric search for binary black hole systems.

\begin{figure*}[ht!]
    \centering
    \includegraphics[trim=10 10 10 10, clip, width=\textwidth]{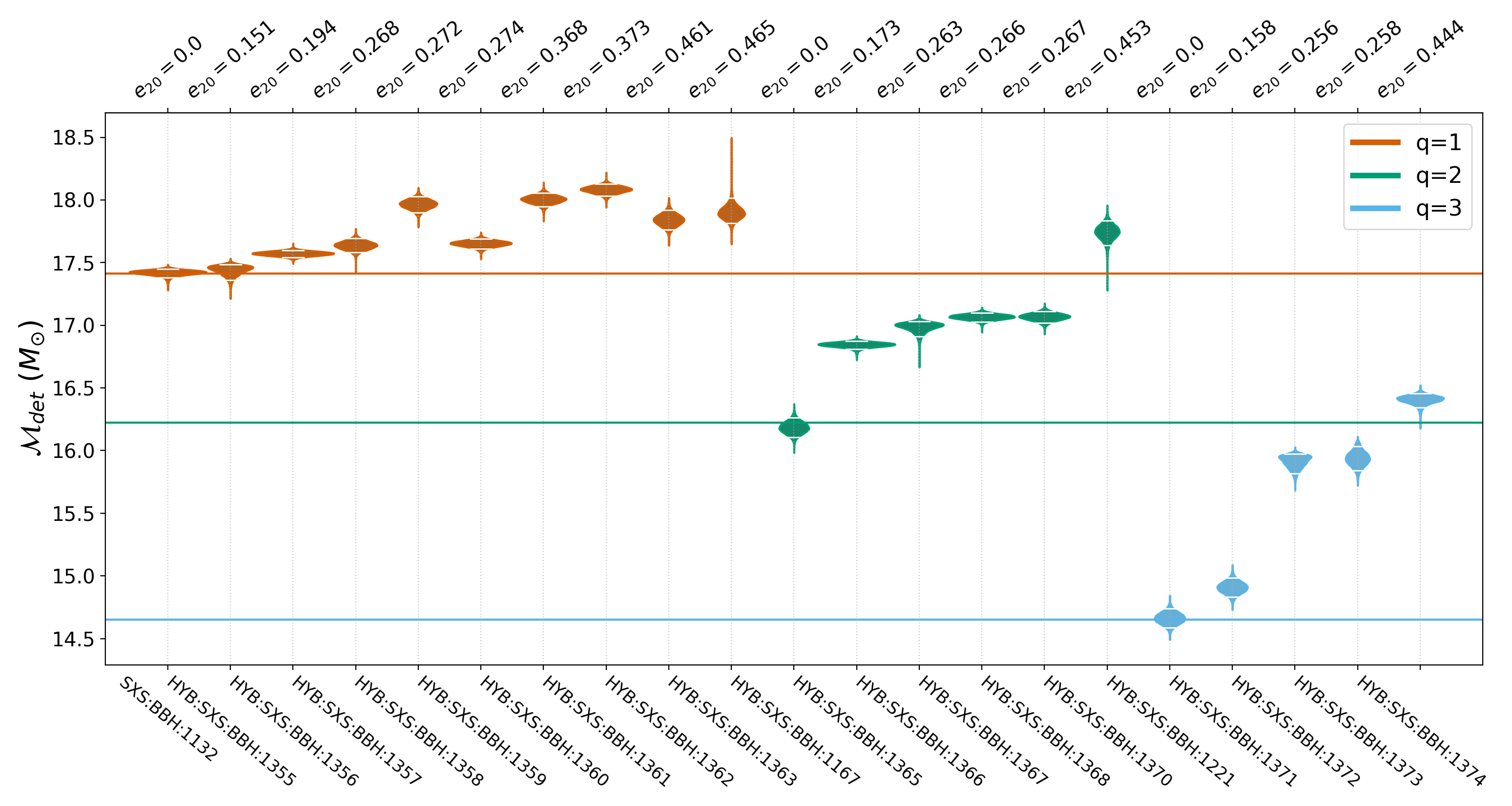}
    \caption{Violin plots showing the chirp mass posteriors for various hybrids. The horizontal axis shows the hybrid ID and corresponding eccentricity values  ($e_{20}$) using \textsc{gw-eccentricity} package~\cite{Shaikh:2023ypz} at the bottom and the top respectively. Different colours correspond to different mass ratios with the coloured horizontal lines denoting the injected values of chirp mass for the respective mass ratios. The small horizontal white lines inside the violins denote the 90\% credible interval. It can be seen that the posteriors for hybrids with low eccentricities include the injected values whereas as the eccentricity is increased, the chirp mass posteriors become biased. The matched filter SNRs lie in the range of $\sim 29 - 45$ depending on the mass ratio and eccentricity values of the system.}
    \label{fig:violin_pe_plot}
\end{figure*}
\subsection{Figures of merit: Overlap distribution }
\label{sec:method-pval}

A waveform consistency test has been developed to quantify this discrepancy \cite{Gayathri:2020coq, Szczepanczyk:2020osv}. This test involves performing a series of dedicated CBC injections with the samples derived from the posterior distributions of the source parameters. In this test, randomized samples are injected (extracted from PE analysis of the event) into the GW data near the event times. Then, we run the cWB waveform reconstruction algorithm on this data. These randomly injected samples are referred to as ``null-sources" injections, while the waveform reconstructed from the actual event data is known as ``on-source". At the end of the cWB analysis, we have waveform reconstruction for each injection.

The waveform match or overlap is defined as,
\begin{equation}
    {\cal{O}}(h_1,h_2)=\frac{\langle h_1|h_2 \rangle}{\sqrt{\langle h_1|h_1 \rangle \langle h_2|h_2\rangle}}
    \label{eq:match}
\end{equation}
where $h_1$ and $h_2$ are the two whitened waveforms, and $\langle\cdot|\cdot\rangle$ represents the noise-weighted inner product, which is defined as,
\begin{equation}
 {\langle h_1|h_2\rangle=\sum_{k=1}^{N} \int_{t_1}^{t_2} h_{1k}(t)h_{2k}(t)dt}   
\end{equation}
where $N$ is the number of detectors in the network and $[t_1,t_2]$ is the time range of the reconstructed signal~\cite{Szczepanczyk:2020osv}. An overlap value of $1$ indicates a perfect overlap between waveforms, while a value close to $0$ indicates that the correlation between waveforms is nil. 

 Let us assume that $W_i$ represents the cWB point waveform reconstruction estimate of $h_i$.
The waveform $W = {W_k(t)}$ is estimated through a model-independent point estimate and a selected whitened waveform from a given signal model $h = {h_k(t)}$, where $k$ represents detector index. We estimate an overlap between the off-source injection ($h_i$) and its cWB waveform reconstruction {($W_i$)}, which we consider as the null-distribution ${\cal{O}}_o(W_i,h_i)$, where index $i$ represents injection. 
Additionally, we compute the on-source overlap distribution (null distribution) denoted as ${\cal{O}}(W,h_i)$ and $W$ is a reconstruction of the event. The null distribution is calculated to understand the cWB reconstruction errors. Note also that a comparison with a null distribution is integral in confirming whether or not the signal carries additional physics (in our case eccentricity) absent in the recovery model (here a circular model) used for PE; while both distributions suffer from reconstruction errors, only on-source distribution is sensitive to the additional physics.\footnote{Further, the reconstruction errors may be sensitive to variations in the values for signal parameters. For instance, we obtain the on-source distribution by comparing the reconstructed signal with reconstructions of signals drawn from the posterior (statistical representatives of the actual signal) and thus there is possible compounding of reconstruction errors since the cWB pipeline will see each sample waveform differently.}

The on-source overlap, denoted as ${\cal{O}}$, is calculated using the point waveform reconstruction $W$ for the GW event and $h_p$ as a proxy for the true GW signal, selected from the posterior distribution. If the model accurately describes the GW event, we anticipate that the overlap ${\cal{O}}(W, h_i)$ will fall within the null distribution. 
If the observed signal $W$ contains additional information, such as eccentricity, that is not accounted for by the waveform model used for PE, the on-source overlap ${\cal{O}}$ will decrease and deviate from the null distribution. These deviations give a hint of the presence of eccentricity, which we can use to subselect interesting GW candidates.  

We define the equation for the figure of merit as follows,
\begin{equation}
\Delta_{\rm median} = \rm{M_1-M_2}
\label{eq:merit}    
\end{equation}
where $\rm{M_1}$ and $\rm{M_2}$ are the medians of null and on-source distribution respectively.
Since the two histograms are obtained independently, related errors are not correlated to each other and the errors in $\Delta_{\rm{median}}$ may be computed using the formula,
\begin{equation}
\sigma=\sqrt{\sigma_{M_1}^2 + \sigma_{M_2}^2}
    \label{eq:error}
\end{equation}
 where $\sigma_{M_1}$ and $\sigma_{M_2}$ are the 90\% credible intervals for null and on-source distribution respectively.

\section{Results}
\label{sec:res}

\subsection{Parameter estimation}
\label{sec:res-inj}

In this section, we perform injection studies using the eccentric hybrids constructed in Chattaraj et al. \cite{Chattaraj:2022tay} to assess the biases that are introduced when quasi-circular waveforms are used to recover eccentric signals. The complete information of the hybrids used for injections is given in Table I in~\cite{Chattaraj:2022tay}. We inject the signals in Gaussian noise simulated from the power spectral density of the detectors and use the inspiral-merger-ringdown waveform model \textsc{IMRPhenomXAS} \citep{Pratten:2020fqn} for recovery. We employ Bayesian inference to compute the likelihood and, hence, construct the posteriors. We assume that our sources are at a distance of $410$~Mpc and inclined at an arbitrary angle of $30^\circ$ to the line of sight. We have fixed the total mass to 40$\rm{M_\odot}$ and the mass ratio varies as [1,2,3] depending on the hybrid (see Table I in \cite{Chattaraj:2022tay}). The right-ascension ($\alpha$), declination ($\delta$), and polarization ($\psi$) angles are chosen arbitrarily with the values $30^\circ$, $45^\circ$, and $60^\circ$ respectively, and the geocent time ($t_\text{gps}$) is taken to be 1137283217~s. Since the signal-to-noise ratio (SNR) of a GW signal depends not only on intrinsic parameters, such as mass and eccentricity but also on extrinsic parameters, variation in these parameters can result in different SNRs, thereby affecting the widths of the posteriors presented in this analysis.

The Bayesian posterior probability for a parameter $\Vec{\theta}$, given the data $\Vec{s}$ and a GW model $\mathcal{H}$, is given by

\begin{equation}
  p(\Vec{\theta}|\Vec{s},\mathcal{H}) = \frac{\mathcal{L}(\Vec{s}|\Vec{\theta},\mathcal{H}) \mathcal{\pi}(\Vec{\theta},\mathcal{H})}{\mathcal{Z}(\Vec{s})}\,,
\end{equation}
where $\mathcal{L}(\Vec{s}|\Vec{\theta},\mathcal{H})$ represents the likelihood, $\mathcal{\pi}(\Vec{\theta,\mathcal{H}})$ is the prior, and $\mathcal{Z}(\Vec{s}|\mathcal{H})$ represents the evidence. We use \texttt{PyCBC}~\cite{Biwer:2018osg} to create injections, \texttt{LALSimulation} \cite{lalsuite} for generating waveforms, and the \emph{nested} sampling algorithm~\cite{2004AIPC..735..395S} implemented through \texttt{dynesty}~\cite{speagle2020dynesty} sampler in {\tt bilby}~\cite{Ashton:2018jfp} and {\tt bilby\_pipe}~\cite{Romero-Shaw:2020owr} for parameter estimation. 

All injections are nonspinning, so the component spin vectors have been set to zero during recovery. We sample the parameter space that includes chirp mass ($\mathcal{M}$), inverse mass ratio ($q_\text{inv} = m_2/m_1$), geocentric time ($t_c$), luminosity distance ($d_L$), phase angle ($\phi_c$), inclination angle ($\theta_\text{jn}$), right ascension($\alpha$), declination($\delta$), and polarization angle ($\psi$). The complete information on priors is presented in the Appendix~\ref{appendix:priors}. In the following subsections, we use the term ``recovery" to describe a result where the injected value falls within the 90\% credible interval of the posterior, and the systematic bias (defined as the difference between the median value and the injected value) is smaller than the posterior's width at 90\% confidence. We consider a result to exhibit significant bias if the injected value lies entirely outside the 90\% credible interval of the posterior. As noted earlier, these biases depend on the SNRs, which in this study are within the range of typical SNRs observed in gravitational wave event catalogues. We use the three detector network involving two LIGO detectors with design sensitivities of Advanced LIGO \cite{PyCBC-PSD:aLIGO} and one Virgo detector with design sensitivity of Advanced VIRGO \cite{PyCBC-PSD:AdvVirgo} to perform all the parameter estimation analyses
shown here. 

We present the results for parameter estimation in Fig.~\ref{fig:violin_pe_plot}. Here we plot the chirp mass posteriors for all the hybrids in the form of violin plots. The bottom and top axes of the plot label the simulation ID and the eccentricity value at 20 Hz for the corresponding hybrids, respectively. The colours orange, green, and blue correspond to mass ratios 1, 2, and 3, respectively, with the injected chirp mass values shown as the same colour horizontal lines on the plot. Small white lines inside the filled curve denote 90\% credible interval for each violin.

It can be seen that the injected values of chirp mass lie within the 90\% credible intervals for simulations with low values of eccentricity such as SXS:BBH:1132, HYB:SXS:BBH:1355, HYB:SXS:BBH:1167 and HYB:SXS:BBH:1221 but as we go to hybrids with higher eccentricities, the posteriors become increasingly biased. Since an eccentric signal is shorter in length than its circular counterpart, when eccentric injections are analysed using quasicircular waveforms, higher mass templates are picked up to compensate for the shorter signals. Hence we see a bias toward higher chirp mass as the value of eccentricity is increased. For hybrids with the same mass ratio and with similar values of eccentricity (e.g. HYB:SXS:BBH:1358 and HYB:SXS:BBH:1359), a difference in the mean anomaly (see Table I in \cite{Chattaraj:2022tay}) can result in different biases.\footnote{ Note that both the value of eccentricity at a reference frequency and orientation of the binary’s eccentric orbit (characterised by the mean anomaly) will induce modulations in the amplitude and the phase of the signal and thus impacting parameter recovery if waveforms employed in the parameter estimation do not sample in these.  
.} 
In the following sections, we highlight how these biases result in various differences in the reconstructed waveforms and hence help us identify the signature of eccentricity with cWB.

\subsection{cWB waveform reconstruction}
\label{sec:res-cwb}
\subsubsection{Reconstructed waveforms}

We perform the cWB-waveform reconstructions for all injections mentioned in the PE section. The details about cWB reconstruction are discussed in Sec. \ref{sec:method-cwb}. These analyses were performed in Gaussian noise with Advanced LIGO and Virgo sensitivity. 

\begin{figure*}[htp!]
    \centering
    \begin{subfigure}[t]{0.46\linewidth} 
        \centering
        \includegraphics[width=\linewidth]{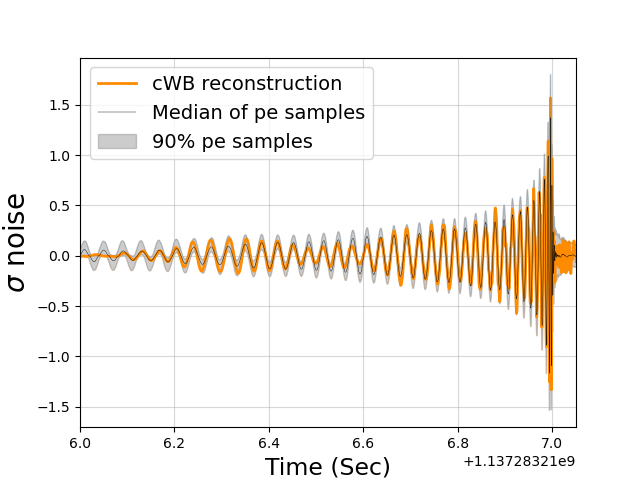}
        \caption{SXS:BBH:1132 $q=1$ and zero-eccentricity }
    \end{subfigure}
    \hfill
    \begin{subfigure}[t]{0.46\linewidth} 
        \centering
        \includegraphics[width=\linewidth]{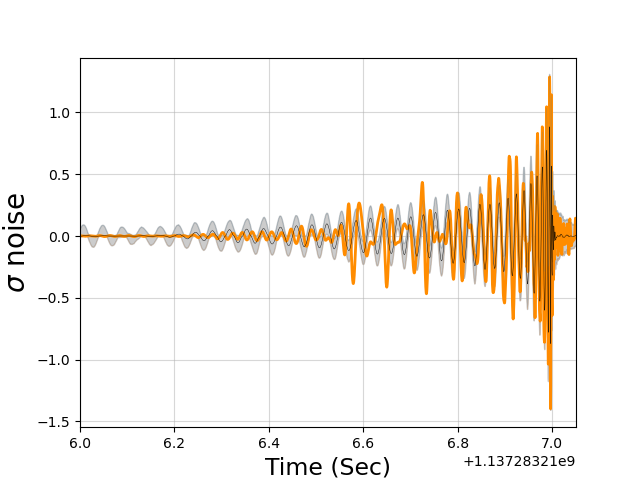}
        \caption{HYB:SXS:BBH:1363 $q=1$ and  $e_{20}=0.465$}
    \end{subfigure}
    \vskip 0.5em 
    \begin{subfigure}[t]{0.46\linewidth}
        \centering
        \includegraphics[width=\linewidth]{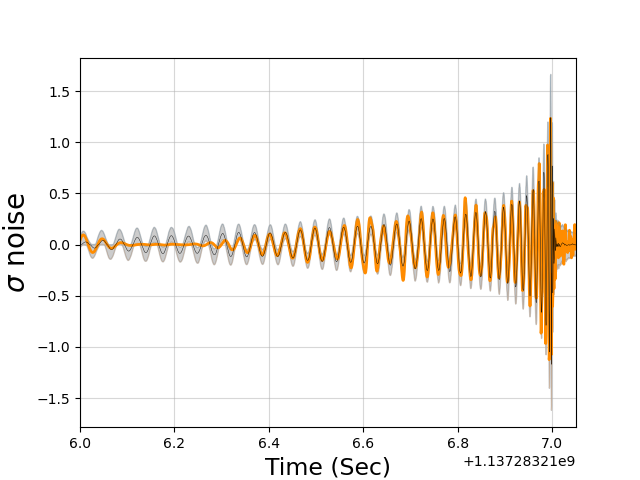}
        \caption{HYB:SXS:BBH:1167 $q=2$ and zero-eccentricity}
    \end{subfigure}
    \hfill
    \begin{subfigure}[t]{0.46\linewidth}
        \centering
        \includegraphics[width=\linewidth]{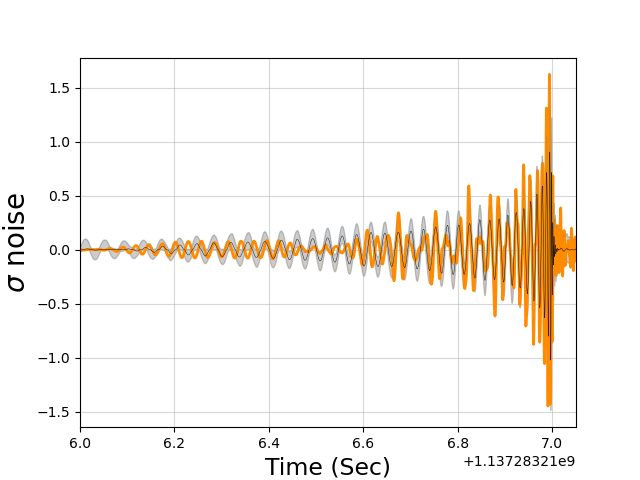}
        \caption{HYB:SXS:BBH:1370 $q=2$ and  $e_{20}=0.453$}
    \end{subfigure}
    \vskip 0.5em
    \begin{subfigure}[t]{0.46\linewidth}
        \centering
        \includegraphics[width=\linewidth]{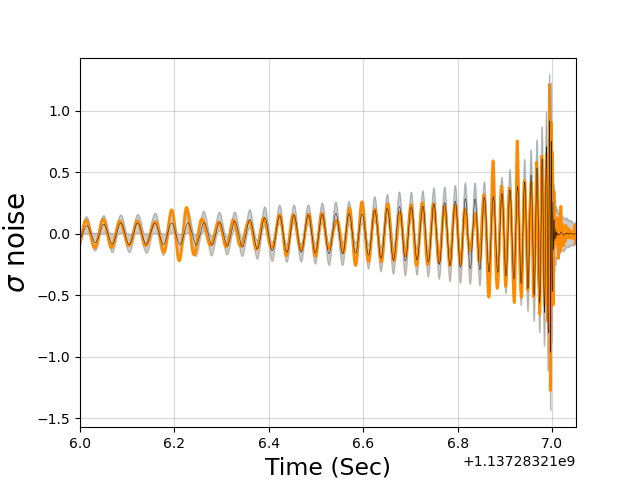}
        \caption{HYB:SXS:BBH:1221 $q=3$ and zero-eccentricity}
    \end{subfigure}
    \hfill
    \begin{subfigure}[t]{0.46\linewidth}
        \centering
        \includegraphics[width=\linewidth]{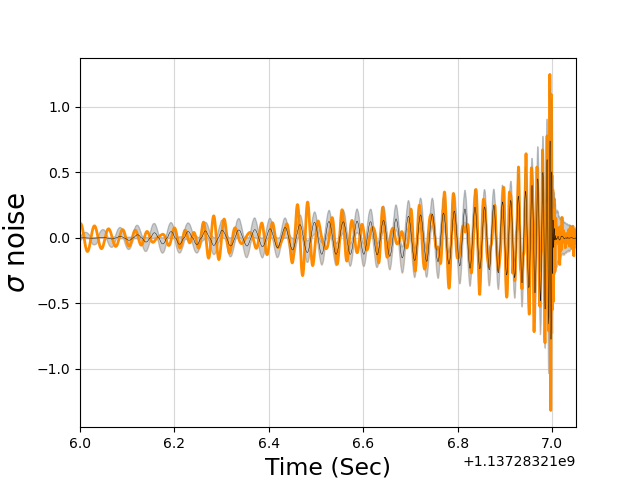}
        \caption{HYB:SXS:BBH:1374 $q=3$ and  $e_{20}=0.444$}
    \end{subfigure}

    \caption{Reconstructed whitened strain using cWB for a given parameter estimation run. The orange curve represents the whitened waveform of the hybrid injection as reconstructed by cWB for the LIGO Livingston detector. The light gray shaded regions denote the 90\% credible intervals derived from the waveform reconstruction of the PE samples. The top, middle, and bottom rows represent results from hybrid analysis for $q=1, 2 ,3$, respectively. The left and right columns denote zero and nonzero eccentricity, respectively.}
    \label{fig:wavform_rec}
\end{figure*}

Figure \ref{fig:wavform_rec} shows the reconstructed whitened strain for parameter estimation of selected hybrids. The orange curve represents the reconstructed whitened waveform for a hybrid injection by cWB for the LIGO Livingston detector which also acts as a reference detector for our analysis. The light gray shaded regions denote the 90\% credible intervals derived from the waveform reconstruction of the PE samples, and the black curve corresponds to the median of these reconstructed waveforms.  The top, middle and bottom rows represent results from selected hybrids with $q=1,2,3$, respectively. The left and right columns refer to zero and maximum eccentricity hybrid injection, respectively, for a given mass ratio. It is well understood that eccentricity influences the signal duration; for a fixed-mass binary, the signal duration decreases as eccentricity increases. This trend is observable in Fig \ref{fig:wavform_rec}, moving from left to right. 

We observe that for hybrids with no eccentricity (left column of Fig.~\ref{fig:wavform_rec}), the reconstructed signal lies within the 90\% credible interval (shaded gray region). However, for eccentric hybrid injections (right column of Fig.~\ref{fig:wavform_rec}), discrepancies arise between the 90\% reconstruction band and the injected reconstruction. These discrepancies become more pronounced as eccentricity increases and are observed in both the time and frequency domains. For hybrids with higher eccentricity (e.g., HYB:SXS:BBH:1361, HYB:SXS:BBH:1363), the reconstructed strain not only falls significantly outside the gray region but also shows a considerable phase deviation from the median of the PE samples.

\begin{figure*}
    \centering
        \centering
    \begin{subfigure}[t]{0.45\linewidth} 
        \centering
        \includegraphics[width=\linewidth]{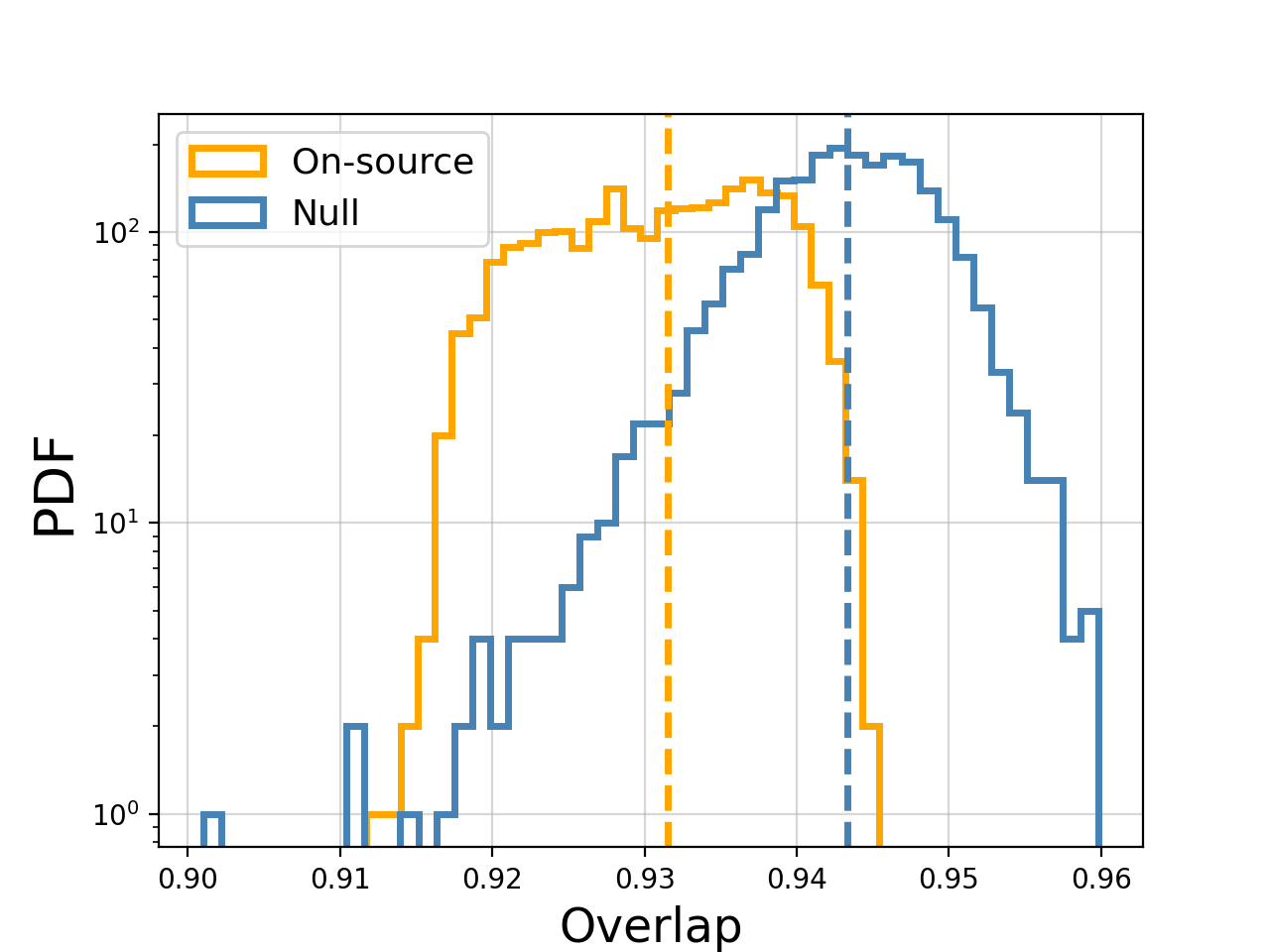}
        \caption{SXS:BBH:1132 $q=1$ and zero-eccentricity }
    \end{subfigure}
    \hfill
    \begin{subfigure}[t]{0.45\linewidth} % Reduced width
        \centering
        \includegraphics[width=\linewidth]{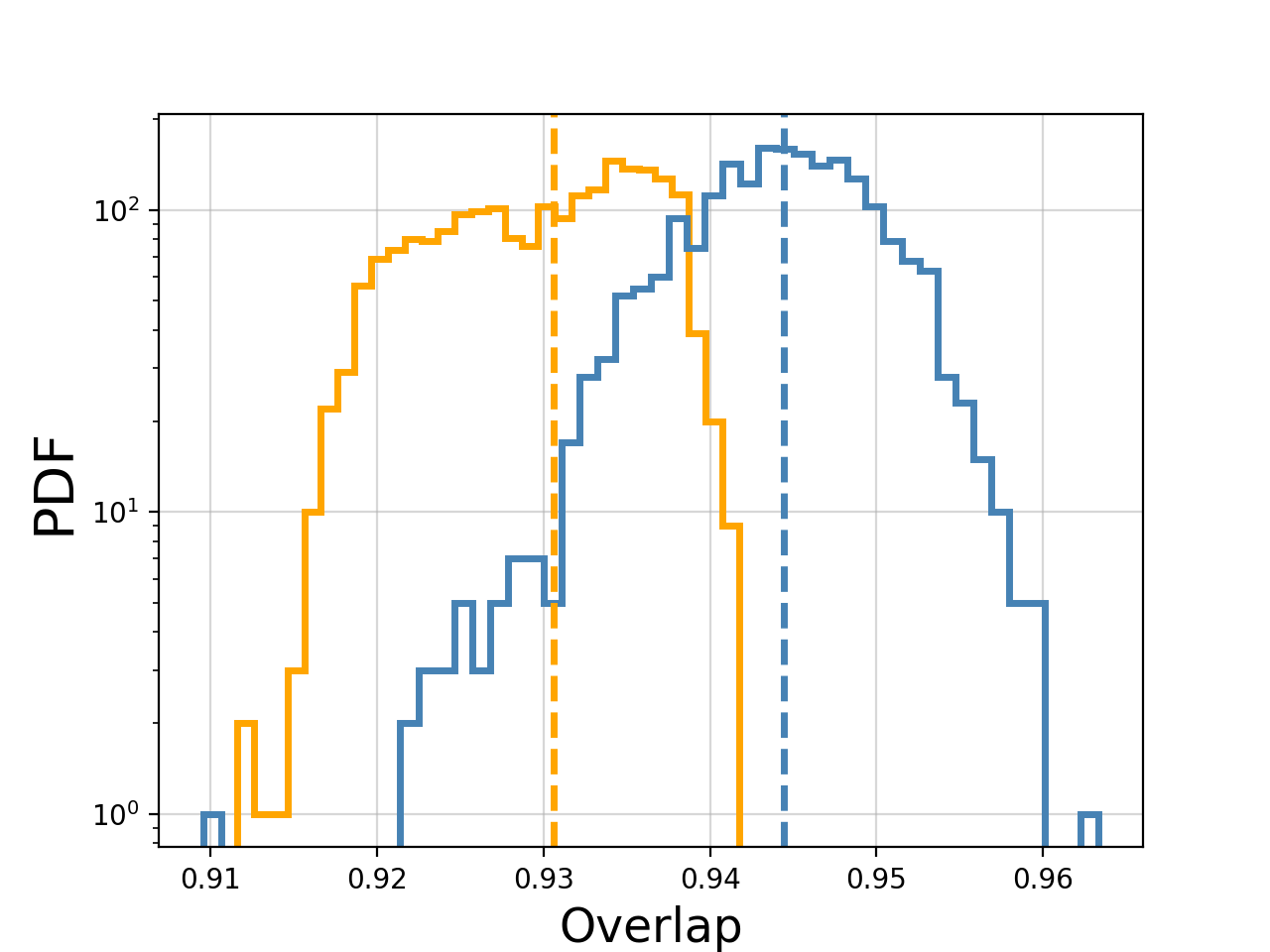}
        \caption{HYB:SXS:BBH:1355 $q=1$ and  $e_{20}=0.194$}
    \end{subfigure}
    \vskip 0.5em 
    \begin{subfigure}[t]{0.45\linewidth}
        \centering
        \includegraphics[width=\linewidth]{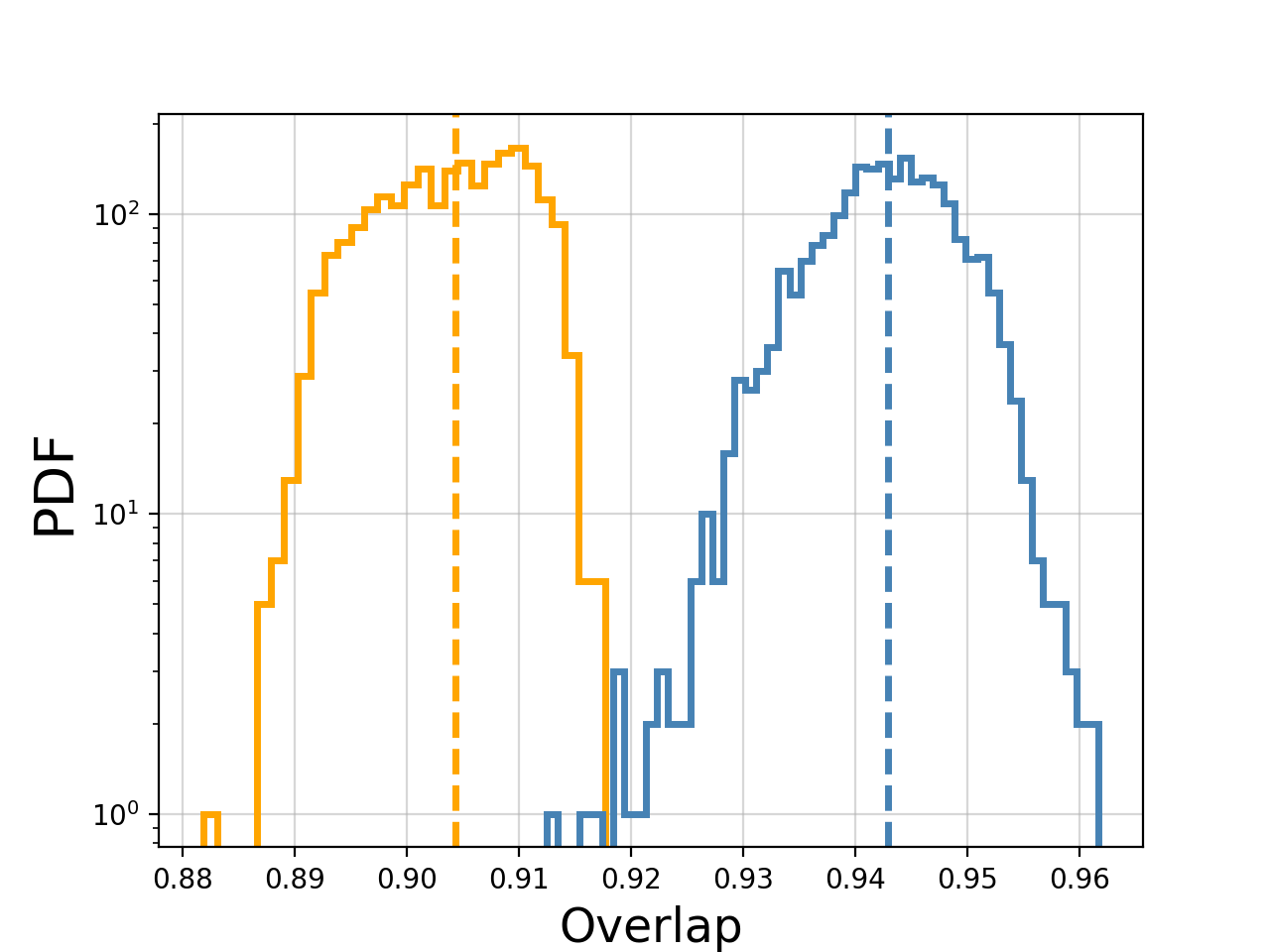}
        \caption{HYB:SXS:BBH:1356 $q=1$ and  $e_{20}=0.268$}
    \end{subfigure}
    \hfill
    \begin{subfigure}[t]{0.45\linewidth}
        \centering
        \includegraphics[width=\linewidth]{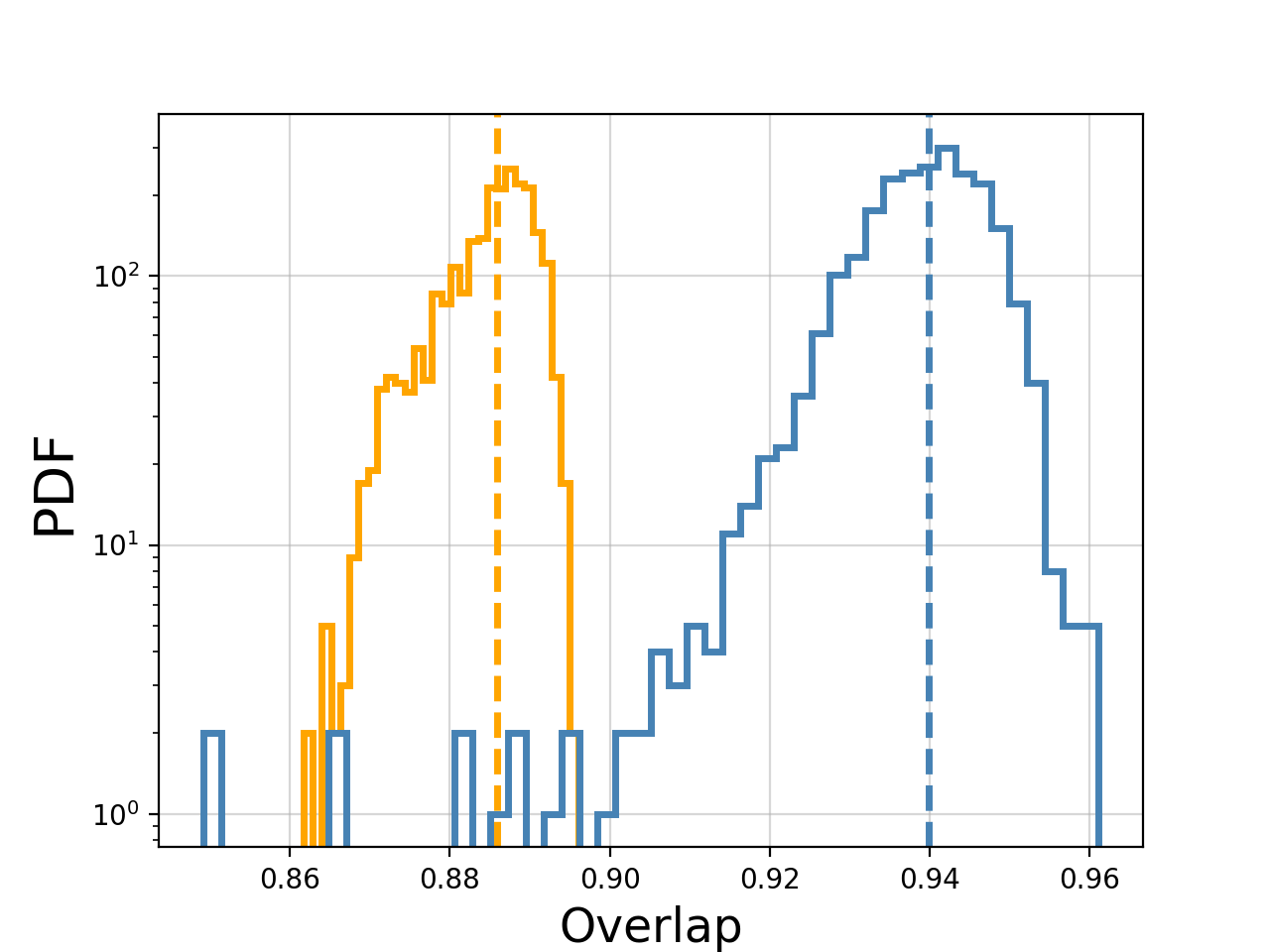}
        \caption{HYB:SXS:BBH:1358 $q=1$ and  $e_{20}=0.272$}
    \end{subfigure}
    \vskip 0.5em
    \begin{subfigure}[t]{0.45\linewidth}
        \centering
        \includegraphics[width=\linewidth]{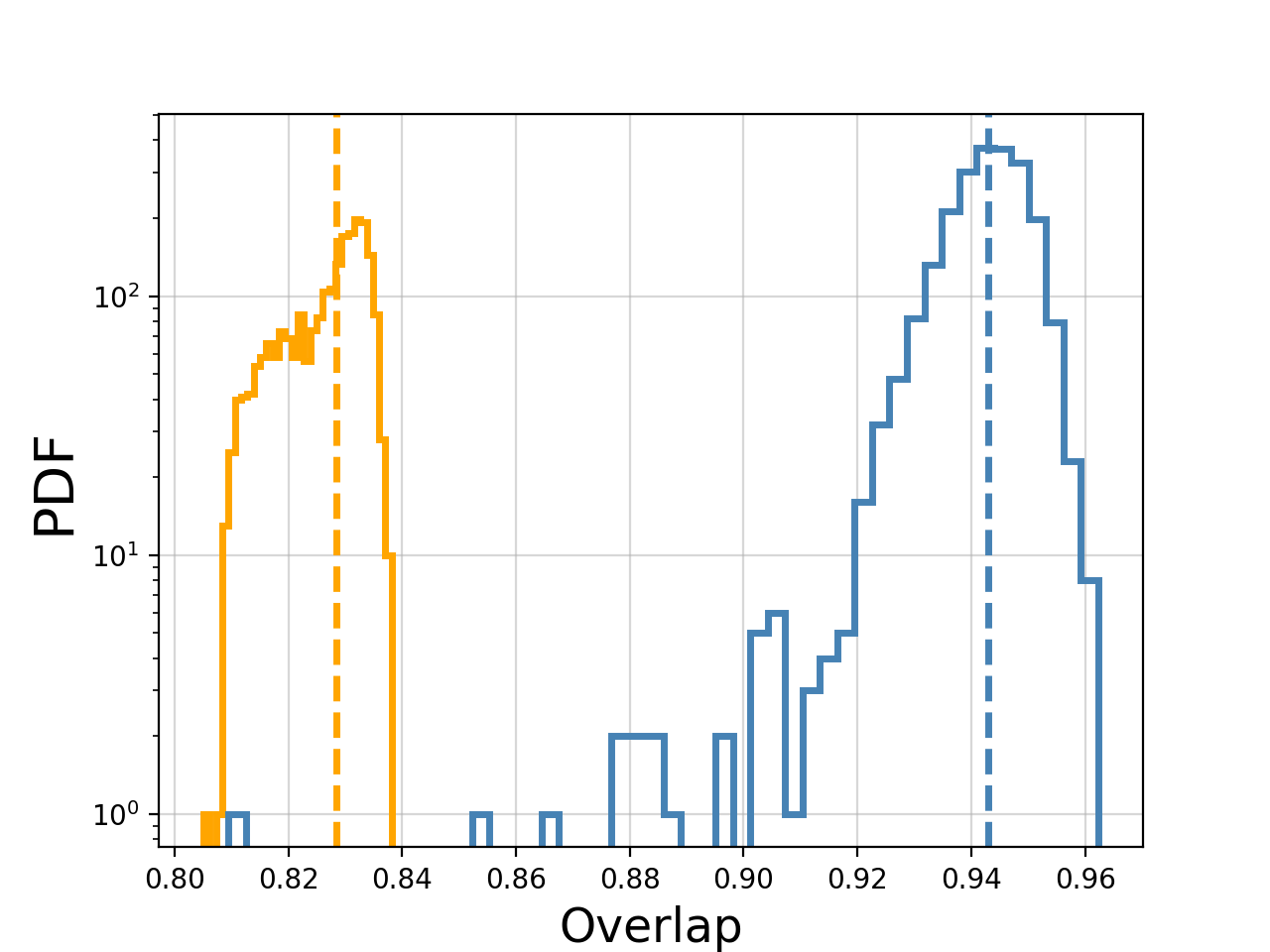}
        \caption{HYB:SXS:BBH:1361 $q=1$ and  $e_{20}=0.373$}
    \end{subfigure}
    \hfill
    \begin{subfigure}[t]{0.45\linewidth}
        \centering
        \includegraphics[width=\linewidth]{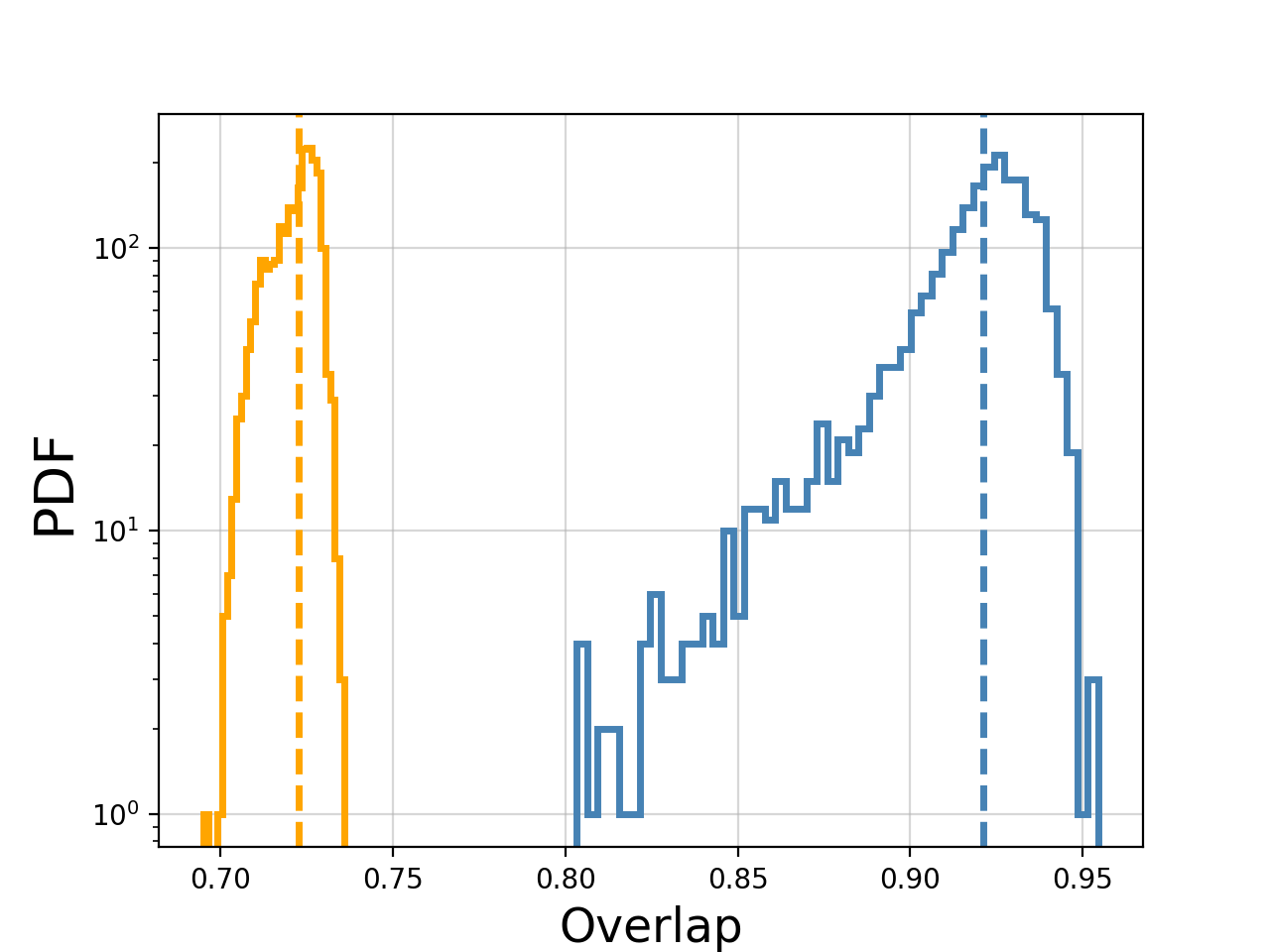}
        \caption{HYB:SXS:BBH:1363 $q=1$ and  $e_{20}=0.465$}
    \end{subfigure}
    
    \caption{Overlap histograms showing on-source (orange) and null (blue) distributions for different cWB reconstruction analyses with $q=1$. The dashed orange and blue lines represent medians for on-source and null distributions, respectively. Subplot (a) corresponds to zero eccentricity, while subplots (b), (c), (d), (e), and (f) represent analyses of injections with increasing eccentricity values.  A similar trend (with increase in eccentricity) is observed for higher mass ratio cases ($q=2, 3$); see also Figure~\ref{fig:median} and discussion around it.}
    \label{fig:match}
\end{figure*}

\subsubsection{Figures of merit: Overlap distribution }
\label{sec:delta_median}

\begin{figure}[h!]
    \centering
        \centering
   
        \centering
        \includegraphics[width=\linewidth]{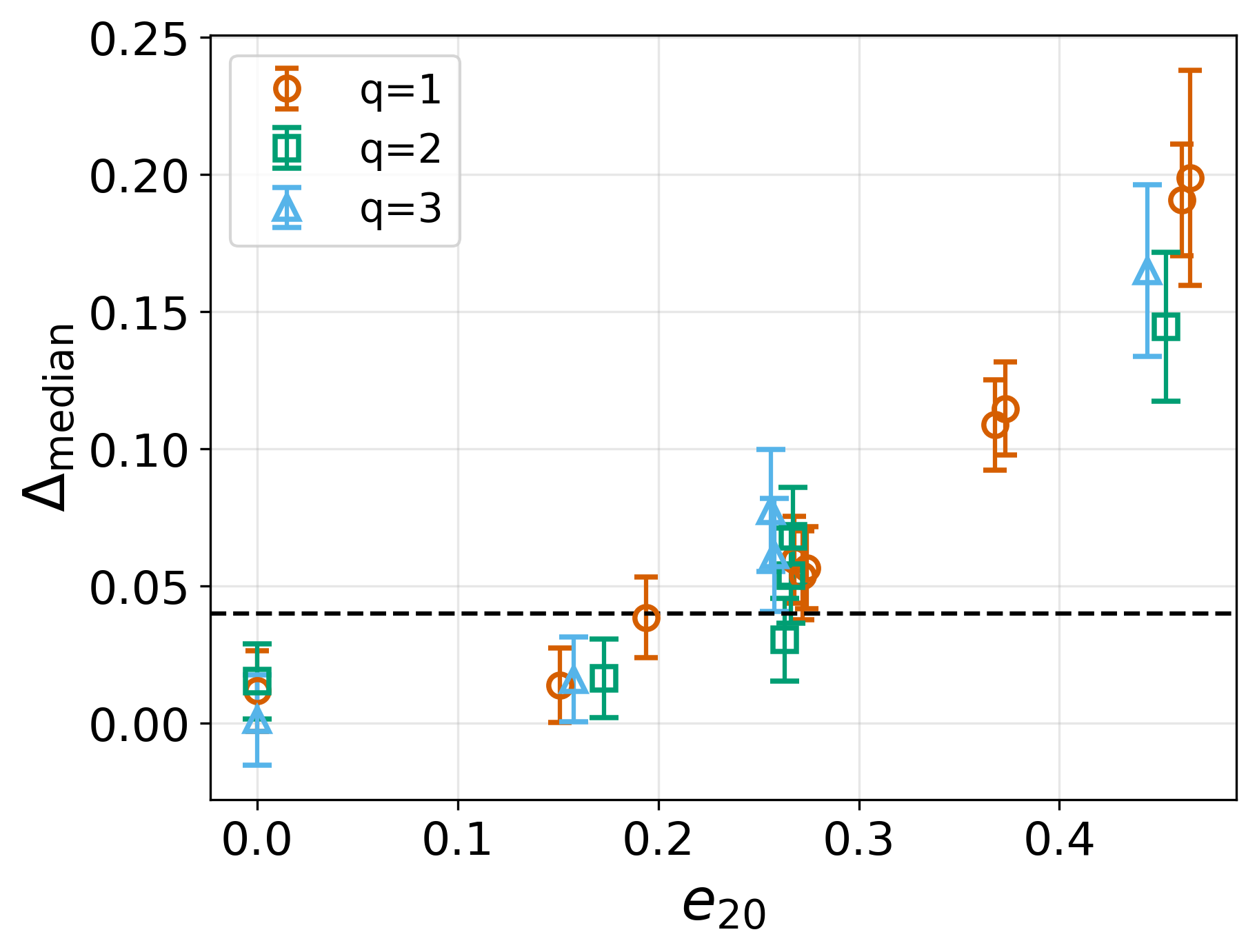}
    
    \hfill
    
    \caption{Plot of ${\Delta_{\text{median}}}$ for different values of eccentricity. Each color corresponds to a specific mass ratio, illustrating the variation across events. The black dashed line represents  our exclusion criterion using an upper bound on $\Delta_{\rm{median}}$ value highlighted in the flowchart of Fig.~\ref{fig:method}.}\label{fig:median}
        \end{figure}

In this section, we present the results of the figure-of-merit analysis conducted for the injected signals. Overlap factors were computed using Eq.~\eqref{eq:match} for all simulated signals, following the methodology outlined in the flow chart. Figure.~\ref{fig:match} illustrates the overlap distributions for on-source (orange) and null (blue) cases across various cWB reconstruction analyses for \(q=1\). The vertical dashed lines denote the median for on-source and null distributions respectively.

Subplot (a) corresponds to zero eccentricity, while subplots (b), (c), (d), (e), and (f) depict analyses of injections with eccentricity arranged in increasing order. The null distribution compares the overlap between the injected PE sample waveform and its cWB waveform reconstruction. In contrast, the on-source distribution compares the cWB reconstruction of the hybrid injection to the PE sample cWB waveform reconstruction.

We observe that the null histogram exhibits a relatively consistent distribution of overlap values across different cases. However, this behaviour changes for high eccentricity (\(e_{20} = 0.44\)) case, most likely due to a loss in low-energy pixels during the reconstruction process. Significant deviations in the  shared region between the two histograms can be identified with increasing eccentricity. Beyond a certain eccentricity threshold, the area of the shared region diminishes to nearly zero, indicating a substantial difference between the PE sample cWB waveform reconstruction and the hybrid injection reconstruction.

To quantify this behaviour, we introduce a new metric that measures the difference between the medians of the two histograms (Eq.~\ref{eq:merit}) as a function of eccentricity. This metric provides a clearer indication of how waveform reconstruction consistency degrades with increasing eccentricity. We present a figure of merit Fig.~\ref{fig:median} for different hybrids. Each data point in the scatter plot corresponds to a specific eccentricity and mass ratio, capturing trends and variations across these parameters. As eccentricity increases, for $q=1$ the value of $\Delta_{\rm{median}}$ increases from 0.01 ($e_{20}=0$) to  0.2 ($e_{20}=0.47$) indicating that the eccentricity feature is observable in $\Delta_{\rm{median}}$ (Fig~\ref{fig:match}). For $q=2,3$ the value of $\Delta_{\rm median}$ increases from 0.02 ($e_{20}=0$) to 0.14 ($e_{20}=0.45$) and from  0.001 ($e_{20}=0$) to 0.17 ($e_{20}=0.44$), respectively. As can be verified from Figure~\ref{fig:median}, a criterion of 4\% for upper bounds on $\Delta_{\rm median}$ values (shown as black-dashed line) eliminates cases with eccentricity values (estimated at 20 Hz for a binary of total mass of $40\rm {M_\odot}$ using the \textsc{gw-eccentricity} package of Ref~\cite{Shaikh:2023ypz}) smaller than 0.17 as potential eccentric sources. As indicated in Figure \ref{fig:method}, we use this to classify the source as a potential eccentric source or otherwise for follow up studies. We have plotted the figure of merit with error bars where, errors in the median shift are calculated using standard error propagation method (Eq.~\ref{eq:error}).
These deviations represent the combined contributions of biases in CBC parameter estimation and effects from cWB waveform reconstruction. It is important to note that this analysis was conducted in Gaussian noise; for real noise scenarios, these results will be influenced by noise fluctuations. 

\section{Conclusion and future directions}
\label{sec:concl}

In this section, we summarize our results from the above analysis. The motivation to hunt for eccentric binary mergers is to gain information about the formation channel of the binaries. A binary formed in dense stellar environment or in a three-body interaction can have residual eccentricity $e_{20}\gtrsim0.1$
which is likely to be detected in the ground-based detectors. Hence, ignoring eccentricity in the waveform model can result in a loss of SNR in a matched filter search. However, computational expenses continue to increase exponentially in parameter estimation studies with added parameters. Therefore, we explore the detectability of eccentricity using unmodelled waveform reconstruction algorithm.

Our results can be divided into two parts: parameter estimation analysis and waveform reconstruction study (see Fig.~\ref{fig:method} for a flowchart describing the methodology). We inject hybrid signals of total mass 40$\rm{M_\odot}$ into Gaussian noise and perform PE using a quasicircular waveform model, \textsc{IMRPhenomXAS}~\citep{Pratten:2020fqn}. The violin plot in Fig.~\ref{fig:violin_pe_plot} shows the effect of ignoring eccentricity greater than 0.17 at 20Hz ($e_{20}$) for $q$= 1,2 and 3 as the recovered chirp mass is outside the 90\% credible interval. Similar biases are also noticed in recovering other intrinsic parameters such as total mass and mass ratio with increasing eccentricity. Additionally, in the waveform reconstruction analysis (see Fig.~\ref{fig:wavform_rec}) we observe that for $e_{20}\gtrsim0.17$ the reconstructed waveform is inconsistent with the 90\% credible interval and is out of phase. We compute the overlap factor for on-source and null distributions discussed in Sec \ref{sec:method-pval}. For the on-source analysis, we find that with increasing eccentricity, the overlap factor reduces sharply and reaches $\sim$ 75\% for $e_{20}\simeq0.44$. As a result, the on-source distribution deviates from the null distribution, which is also evident from Fig~\ref{fig:median}. 
Thus, we can conclude that the orbital eccentricity of a binary system, which is a subdominant effect, can be isolated even in cases where eccentric waveforms are not readily accessible. The methodology is so robust that it can be employed to infer the presence of any sub-dominant physical effect, such as the presence of precession effects in the signal due to spin and/or higher-order modes apart from eccentricity. In future work, we plan to extend the methods presented here to analyse real events, which, in principle, can include a combination of these subdominant effects.\\
\\
Overall, our results demonstrate that eccentricity introduces measurable inconsistencies in waveform reconstruction. These findings emphasize the necessity for incorporating eccentricity into waveform models and analysis pipelines to enhance reconstruction accuracy for gravitational wave signals.

\acknowledgments
We thank Juan Calder\'on Bustillo for his useful comments
on the manuscript.
RD thanks the Gravitation and Cosmology Group at IIT Madras for numerous insightful discussions. R.D. also thanks Tanmaya Mishra for helping fix a few issues with the cWB pipeline installation. 
D.J. acknowledges the Science and Technology Facilities Council (STFC) for support through grants ST/V005618/1 and ST/Y004272/1. G.V. acknowledges the support of NSF under grant PHY-2207728.
S.K. acknowledges the support of NSF under grants PHY 2110060 and PHY-2409372. I.B acknowledges the support of NSF under grants PHY-2309024. 
C.K.M. acknowledges the support of SERB's Core Research Grant No.~CRG/2022/007959. 
Computations were performed on CIT, LHO, and LLO clusters provided by the LIGO Laboratory and supported by the National Science Foundation Grants PHY-0757058 and PHY-0823459. This material is based upon work supported by NSF's LIGO Laboratory, which is a major facility fully funded by the National Science Foundation. We used the following software packages:
{\tt LALSuite}~\cite{lalsuite}, {\tt bilby}~\cite{Ashton:2018jfp}, {\tt bilby\textunderscore pipe}~\cite{Romero-Shaw:2020owr}, \texttt{PyCBC}~\cite{alex_nitz_2020_4134752}, {\tt NumPy}~\cite{2020Natur.585..357H}, {\tt PESummary}~\cite{Hoy:2020vys}, {\tt Matplotlib}~\cite{2007CSE.....9...90H}, {\tt Seaborn}~\cite{Waskom2021}, {\tt jupyter}~\cite{soton403913}, and {\tt dynesty}~\cite{speagle2020dynesty}.
This document has LIGO preprint No. {\tt LIGO-P2400590}.

\appendix
\section{PRIORS USED FOR PARAMETER ESTIMATION}
\label{appendix:priors}

The sample space for parameter estimation includes the following parameters: inverse mass ratio ($q_\text{inv} = m_2/m_1$),\footnote{We use the word \textit{inverse} here to indicate that this is inverse of the mass ratio we have used throughout the paper. Please note that \texttt{bilby} uses the term \texttt{mass\_ratio} for this.} chirp mass ($\mathcal{M}$), luminosity distance ($d_L$), inclination angle ($\theta_\text{jn}$), geocentric time ($t_c$), phase angle ($\phi_c$), right ascension ($\alpha$), declination ($\delta$), and polarization angle ($\psi$). We have put constraint on component masses as [5,50]$M_\odot$. The priors for all the parameters are given in Table \ref{table:priors}.

\begin{table}[ht]
\def\arraystretch{1.3}
\begin{tabular}{|c|c|c|}
\hline
\textbf{Parameter} & \textbf{Prior} & \textbf{Range} \\ \hline
$\mathcal{M}$ & Uniform & $5 \text{ - } 50 M_\odot$ \\ \hline
$q_\text{inv}$ & Uniform & $0.125 \text{ - } 1$ \\ \hline
$d_L$ & \begin{tabular}[c]{@{}c@{}}Uniform \\ Source Frame\end{tabular} & $100 \text{ - } 1000$ Mpc \\ \hline
$\theta_\text{jn}$ & Uniform sine & $0 \text{ - } \pi$ \\ \hline
$\phi_c$ & Uniform & $0 \text{ - } 2\pi$ \\ \hline
$\alpha$ & Uniform & $0 \text{ - } 2\pi$ \\ \hline
$\delta$ & Uniform cos & $-\pi/2 \text{ - } \pi/2$ \\ \hline
$\psi$ & Uniform & $0 \text{ - } \pi$ \\ \hline
$t_c$ & Uniform & $t_\text{gps} \pm 0.1$ s \\ \hline
\end{tabular}
\caption{Priors for parameters used in precessing spin recoveries.}
\label{table:priors}
\end{table}
\clearpage
\bibliographystyle{apsrev4-1}
\bibliography{master_updated}
\end{document}